\documentclass[fleqn,twoside]{article}
\usepackage{amsmath}
\usepackage{espcrc2}

\usepackage{graphicx}
\usepackage{psfrag}
\usepackage{booktabs}

\title{
{
\vspace{-3.0cm} \normalsize \hfill
\parbox{30mm}{SFB/CCP-03-36}
}\\[15mm]
Comparison of analytic and numerical results in the XY-model\thanks{presented by T. Korzec}
}

\author{Janos Balog\address[RMKI]{Research Institute for Particle and Nuclear Physics\\ 1525 Budapest 114, Pf. 49, Hungary}, 
       Francesco Knechtli\address[HU]{Institut f\"ur Physik, Humboldt Universit\"at\\ Newtonstr. 15\\ 12489 Berlin, Germany},
       Tomasz Korzec\addressmark[HU],
       and Ulli Wolff\addressmark[HU]}

\begin{document}

\begin{abstract}
We study the two dimensional XY-model with high precision Monte Carlo techniques
and investigate the continuum approach of the step-scaling function 
of its finite volume mass gap. 
The continuum extrapolated results are found consistent
with analytic predictions for the finite volume energy spectrum based on the 
equivalence with sine-Gordon theory.
To come to this conclusion it was essential to use an also predicted
form of logarithmic decay of lattice artifacts for the extrapolation.
\vspace{1pc}
\end{abstract}

\maketitle

\section{Introduction}
\label{intro}
   Interacting quantum field theories in four space time dimensions are
   in all relevant cases impossible to solve exactly and
   results are 
   obtained only in suitable approximation schemes. The situation is different 
   with two dimensional systems which often may 
   be (partially) solved exactly. Such a system is for instance
   the sine-Gordon model for which detailed predictions for
   the finite volume energy spectrum exist \cite{DdV92,DdV97,Feverati98_2}. These predictions
   can  be extended to the two dimensional $O(2)$ non-linear $\sigma$-model
   which is believed to lie in the same universality class as the sine-Gordon
   model at a special value of its coupling. 
   For the derivation of these exact results one adopts some not rigorously
   provable conjectures at intermediate steps, therefore a numerical
   confirmation is desirable. 
         
   We study the massive phase of the XY-model with Monte Carlo techniques and 
   extract the finite volume mass gap from a time-slice (zero momentum)
   correlation function.         
   The continuum extrapolation is performed
   according to a prediction by J.~Balog \cite{Balog2000}. In the continuum limit some
   points of the step-scaling function of the L\"uscher-Weisz-Wolff (LWW) 
   coupling \cite{LWW91} are compared to values obtained from solutions of the 
   Destri-de Vega (DdV) equation.   
   A more detailed account on the subject is given in \cite{Korzec2003_2}.

\section{Sine-Gordon theory in finite volume}
\label{DdV}   
   A complete description of the exact spectrum of the sine-Gordon
   model in finite volume is provided by the DdV non-linear
   integral equations. Following \cite{Ravanini2001} we solve the equations
   iteratively and refer to \cite{Korzec2003_2} for further details of this procedure. 
   We are interested in the point where the sine-Gordon
   model and the $O(2)$ non-linear $\sigma$-model coincide, that is at
   $\beta_{\rm sG} = \sqrt{8\pi}$. For a chosen volume $ML$ 
   (expressed in units the infinite volume mass gap $M$) we
   solve the equations for the ground state and for the 
   first excited state in order to calculate the finite 
   volume mass gap
   \begin{equation}
      M(L) = E_1(L) - E_0(L)
   \end{equation}
   and hence the LWW coupling   
   \begin{equation}
      \bar g^2 = 2M(L)L \, .
   \end{equation}
   At the doubled volume a point of the step scaling function
   \begin{equation} \label{cssf}
      \sigma(2,\bar g^2) = \bar g^2(2L)
   \end{equation}
   is obtained.
   A number of such results is given in the second column of table \ref{stepscaleResults1}.

\section{Numerical work}
\label{montecarlo}
   We simulate the two dimensional XY-model with 
   standard action 
   \begin{equation}
      S = -\beta \sum_{\langle k,l\rangle} \vec s_k \cdot \vec s_l \, .
   \end{equation}
   on $L/a \times 5L/a$ lattices. 
   We apply periodic boundary conditions in the 
   spatial direction and free boundary conditions 
   in the temporal direction. As Monte Carlo algorithm
   we choose the highly efficient single cluster algorithm \cite{Wolff89}
   which does practically not suffer from critical slowing down.
   To measure the time slice correlation function
   \begin{equation}
      G(\tau) = \langle \vec S(t)\cdot \vec S(t+\tau) \rangle, 
      \quad \vec S(t)= \frac{1}{L} \sum_x \vec s(x,t) \, ,
   \end{equation} 
   we  employ Hasenbusch's improved estimator \cite{Hasenbusch94_2}.
   From the correlation function we extract the finite volume
   mass gap
   \begin{equation}
      M(L)a \overset{\tau \to \infty}{=} \ln\left[ \frac{G(\tau)}{G(\tau+1)}\right]
   \end{equation}
   and obtain a value of the LWW-coupling $\bar g^2(L)$.
   To get one point of the lattice step-scaling function
   \begin{equation}
      \Sigma(2,\bar g^2(L),a/L) =  \bar g^2(2L)
   \end{equation}
   we keep $\beta$ fixed and measure the LWW coupling 
   on a lattice of doubled size. The procedure is 
   repeated several times for the same value of the coupling but different
   lattice resolutions $a/L$. An
   extrapolation to $a/L = 0$ yields one point of the 
   continuum step-scaling function that can be compared 
   to the predicted value (\ref{cssf}). For that purpose 
   the numerical value of $ML$ has to be determined first, at which  the
   DdV equations produce the same coupling.

   For models with a Kosterlitz-Thouless phase transition 
\cite{KosterlitzThouless73,Kosterlitz74}
   like the XY-model,
   J. Balog has predicted the leading lattice
   artifacts to be universal and to decay very slowly, i.e.
   proportional to inverse powers of the logarithm 
   of the infinite volume correlation length. Lattice artifacts of the 
   step scaling function have the form
   \begin{equation}
      \Sigma(2, \bar g^2, a/L ) = 
     \sigma(2,\bar g^2) + \frac{c}{(\ln \xi + U)^2} + \ldots \, ,
   \end{equation}
   where $\xi$ is the infinite volume correlation length, $c$ is universal and 
   can be calculated for each volume and $U$ is a non-universal constant 
   ($U = 1.3(1)$ for the standard action \cite{Balog2002}). Corrections to
   this formula are of order $\ln(\xi)^{-4}$.

\section{Results}
\label{results}
\begin{figure} [htb]
      \psfrag{tag1}{$(1.3+\ln \xi)^{-2}$}
      \psfrag{tag2}{$\Sigma(2,\bar g^2 = 1.7865, a/L)$}
      \psfrag{tag3}{MC-data}
      \psfrag{tag4}{solution of DdV eq.}
      \psfrag{tag5}{linear fit}
      \includegraphics [width = \linewidth] {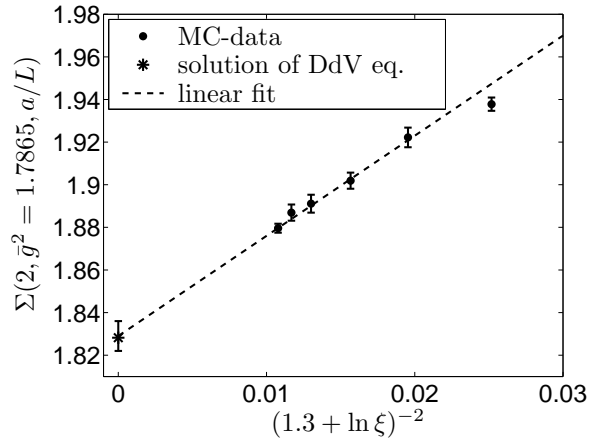}
      \caption{Comparison of MC-data with a numerical solution of the DdV equation at
               $\bar g^2 = 1.7865$.
               The spatial extents $L/a$ of the lattices were $10,\ 20,\ 40,\ 80,\ 120$ and $160$.
               The smallest lattice was discarded in the fit.}
      \label{MCvsDdVFig1}
\end{figure}

   We have performed our calculations at four different values of the LWW coupling.   
   Table \ref{stepscaleResults1} summarizes our results. The lattice artifacts
   prediction includes information about the constant $c$ which is also listed in 
   the table. Fig. \ref{MCvsDdVFig1}
   corresponds to the second line of the table, plots
   for the other points look similar. The theoretical predictions are compatible
   with continuum extrapolated lattice results. The small differences between $c_{\rm th}$
   and $c_{\rm MC}$ may be explained by subleading cutoff effects in the Monte Carlo data.
   The knowledge of the form of lattice artifacts 
   was essential to obtain this result to the precision that is reached.

   \begin{table*}[htb]
      \caption{At different values of the LWW-coupling theoretical predictions (DdV) 
      for the step-scaling
      function are compared with numerical results (MC). Also the slope of the fit $c$ 
      as predicted by
      theory (th) is compared to the numerical value. The last column lists the $\chi^2$ values
      of the extrapolations.\vspace{0.2cm}      
      } \label{stepscaleResults1}
      \begin{tabular}{c c c  c c  c}     
         \toprule
         $ \bar g^2 $&$ \sigma_{\rm DdV}(2,\bar g^2) $&$ \sigma_{\rm MC}(2,\bar g^2)$&$  c_{\rm th} $&$  c_{\rm MC} $&$ \chi^2/_{\rm dof} $\\
	 \midrule
	 $ 3.0038   $&$  4.3895  $&$ 4.40\phantom{0} \pm 0.02\phantom{0}$&$ 2.6176 \pm 0.0002 $&$  2.4 \pm 0.6 $&$   2.51 / 3    $\\
	 $ 1.7865   $&$  1.8282  $&$ 1.829 \pm 0.007                    $&$ 5.30\phantom{00} \pm 0.01\phantom{00} $&$  4.8 \pm 0.5 $&$   0.73 / 3    $\\
	 $ 1.6464   $&$  1.6515  $&$ 1.657 \pm 0.003                    $&$ 5.4\phantom{000} \pm 0.2\phantom{000} $&$  4.3 \pm 0.3 $&$   0.35 / 3    $\\
	 $ 1.6020   $&$  1.6029  $&$ 1.608 \pm 0.004                    $&$ 5.5\phantom{000} \pm 1.5\phantom{000} $&$  4.4 \pm 0.5 $&$   0.90 / 3    $\\  
	 \bottomrule
      \end{tabular}      
   \end{table*}

\section{Conclusions}
\label{conclusions}
   We have investigated the massive scaling limit of the XY-model by means of the 
   step-scaling function of the LWW coupling and found
   it to be consistent with continuum predictions based on the equivalence with sine-Gordon
   theory. 
   A predicted form of lattice artifacts  was also confirmed and has been essential 
   to find the agreement. 
   An extrapolation with powers of $a$ would have led to a significantly different
   continuum result in spite of a perfectly reasonable looking fit. 
   We would like to recall here that deviations from this Symanzik behavior have 
   also been 
   found in the 
   asymptotically free $O(3)$ non-linear $\sigma-$model where they have 
   not yet been understood~\cite{Hasenfratz2000,Hasenbusch2001}.
   
   Our results fit well into the picture that was drawn in \cite{Balog2002},
   where among other things the renormalized 4-point coupling and 2-point correlation functions 
   in the continuum sine-Gordon model were compared to their lattice counterparts in the
   XY-model.

\bibliographystyle{prsty}
\bibliography{../xy_paper/references}







\end{document}